\documentclass[12pt]{article}
\usepackage{times}
\usepackage{geometry}
\usepackage{authblk}
\usepackage[utf8]{inputenc}
\usepackage{wrapfig}
\usepackage[font=small,labelfont=bf]{caption}
\usepackage{enumitem,amsmath,amssymb,subfigure}
\usepackage{ragged2e}
\newlist{thematic}{itemize}{8}
\setlist[thematic]{label=$\square$}
\usepackage{pifont}

\usepackage{natbib}
\usepackage{multicol}[']
\usepackage{graphicx}

\usepackage{indentfirst}

\begin{document}

\begin{titlepage}
\raggedright

\huge
%Planetary Science \& Astrobiology White Paper \linebreak

{Frontiers in Planetary Rings Science}  \linebreak
\normalsize

\textbf{Principal Authors:}
\vspace{-5pt}
 \begin{multicols}{2}
Name: \textbf{Shawn M. Brooks}	
 \linebreak						
Institution: Jet Propulsion Laboratory, California Institute of Technology
 \linebreak
Phone: +1 (818) 393-6380
 \linebreak
Email: Shawn.M.Brooks@jpl.nasa.gov
 \linebreak

 Name: \textbf{Tracy M. Becker}
 \linebreak						
Institution:  Southwest Research Institute
 \linebreak
Email: Tracy.Becker@swri.org
 \linebreak

\end{multicols}
 
 \vspace{-5pt}
\textbf{Co-authors:}
\vspace{-8pt}
  \begin{multicols}{3}
Kevin Bailli\'{e}$^1$  \\
Heidi N. Becker$^2$ \\
E. Todd Bradley$^3$ \\
Joshua E. Colwell$^3$  \\
Jeffrey N. Cuzzi$^4$ \\
Imke de Pater$^5$ \\
Stephanie Eckert$^3$\\
Maryame El Moutamid$^6$ \\
Scott G. Edgington$^2$ \\
Paul R. Estrada$^4$  \\
Michael W. Evans$^7$ \\
Alberto Flandes$^8$ \\
Richard G. French$^9$ \\
Ángel García$^8$ \\
Mitchell K. Gordon$^7$\\
Matthew M. Hedman$^{10}$\\
H.-W. Sean Hsu$^{11}$\\
Richard G. Jerousek$^{12}$\\
Essam A. Marouf$^{13}$  \\
Bonnie K. Meinke$^{14}$ \\
Philip D. Nicholson$^6$\\
Stuart H. Pilorz$^7$ \\
Mark R. Showalter$^7$\\
Linda J. Spilker$^2$ \\
Henry B. Throop$^{15}$\\
Matthew S. Tiscareno$^7$ \\

\end{multicols}

\textbf{Co-signers:}
\vspace{-8pt}
 \begin{multicols}{3}

Jennifer G. Blank$^{4,16}$ \\
Richard J. Cartwright$^7$ \\
Corey J. Cochrane$^2$  \\ 
Luke Dones$^{17} $\\
Cécile C. Ferrari$^{18}$ \\
Robert S. French$^7$ \\
Mark D. Hofstadter$^2$ \\
Sona Hosseini$^2$  \\
Kelly E. Miller$^{17}$ \\
Edgard G. Rivera-Valentín$^{19}$ \\
Abigail Rymer$^{20}$\\
Marshall J. Styczinski$^{21}$  \\
Lynnae C. Quick$^{22}$ \\
Padma Yanamandra-Fisher$^{23}$

\end{multicols}

\begin{figure}[h]
  \begin{center}
  \includegraphics[width=0.8 \textwidth]{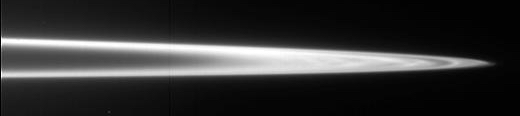}

\end{center}
\end{figure}

\small{ $^1$CNRS, Institut de Mécanique Céleste et de Calcul des Ephémérides; $^2$Jet Propulsion Laboratory, California Institute of Technology; $^3$Univ. of Central Florida; $^4$NASA Ames Research Center ;$^5$Univ. of California, Berkley; $^6$Cornell Univ.; $^7$SETI Institute; $^8$Univ. Nacional Autónoma de México; $^9$Wellesley College; $^{10}$Univ. of Idaho; $^{11}$Univ. of Colorado; $^{12}$Florida Space Institute; $^{13}$San Jose State Univ.; $^{14}$Ball Aerospace; $^{15}$Arctic Slope Technical Services; $^{16}$Blue Marble Space; $^{17}$Southwest Research Institute; $^{18}$Univ. de Paris;  $^{19}$Lunar \& Planetary Institute (USRA); $^{20}$Johns Hopkins, APL; $^{21}$Univ. of Washington; $^{22}$NASA Goddard Space Flight Center; $^{23}$Space Science Institute   }

\normalsize

\begin{center}
\textit{\small A portion of this research was carried out at the Jet Propulsion Laboratory, California Institute of Technology, under a contract with the National Aeronautics and Space Administration (80NM0018D0004).  Copyright: © 2020. All rights reserved.}
\end{center}

%\textbf{Abstract  (optional):}
\end{titlepage}
\raggedright

\pagebreak
\section*{\large Executive Summary}
\vspace{-10pt}
\hspace{1cm} We now know that the outer solar system is host to at least six diverse planetary ring systems, \textbf{each of which is a scientifically compelling target with the potential to inform us about the evolution, history and even the internal structure of the body it adorns.} The Cassini-Huygens mission to Saturn provided a wealth of new information about Saturn’s rings and continues to motivate new work. Likewise, observations from spacecraft and Earth-based observatories have raised many outstanding questions about our solar system's planetary ring systems. Evidence for the formation and ongoing evolution of the planets and their environments can be found in ring structure, composition, and variations caused by magnetospheres, satellites, and planetary internal modes. Rings are more common than once believed; exciting discoveries of ring systems in unexpected environments, including around small bodies like Centaurs and KBOs, compel us to search for new ones. These diverse ring systems represent a set of distinct local laboratories for understanding the physics and dynamics of planetary disks, with applications reaching beyond our Solar System. We highlight the current status of planetary rings science and the open questions before the community to promote continued Earth-based and spacecraft-based investigations into planetary rings. As future spacecraft missions are launched and more powerful telescopes come online in the decades to come, \textit{we urge NASA for continued support of investigations that advance our understanding of planetary rings, through research and analysis of data from existing facilities, more laboratory work and specific attention to strong rings science goals during future mission selections}. We also encourage the active promotion of a diverse, equitable, inclusive and accessible environment in the planetary science community.

\vspace{-10pt}
\section*{\large Giant Planet Ring Systems}
\vspace{-10pt}
\textbf{\textit{Saturn}: }
The popularity of Saturn’s rings stems undoubtedly from their stunning appearance, which led to their early discovery. Christiaan Huygens’ realization that rings encircle Saturn \citep{Huygens} was part of a paradigm shift away from the belief in perfectly spherical heavenly bodies occupying the concordantly perfect heavens \citep{Vanhelden1980}. Saturn’s rings were the subject of repeated investigations in the following centuries. But since visits by Pioneers 10 and 11, Voyagers 1 and 2 and most recently, Cassini, our knowledge has grown vastly. Voyager observations laid the groundwork for Cassini investigations and continue to provide new results \citep{Showalter1991, Spilker2004}.

\hspace{1cm} Cassini orbited Saturn 294 times during 13+ years, nearly half of a Saturnian year. The leap in our knowledge of Saturn’s rings from Cassini is so significant and fundamental that it simply cannot be overstated. Even a basic measure used to characterize rings, optical depth, was discovered to be far more nuanced than our pre-Cassini understanding \citep{Colwell2006, Colwell2007, Hedman2007a}. We cannot summarize all of the new knowledge or review all of the new questions. Instead, we discuss a few significant open questions in order to highlight the depth of the current state of knowledge, as well as the gaps in understanding and avenues for future studies. For a comprehensive summary, see the Rings discipline science section of Volume 1 of the Cassini Mission Final Report\footnote{https://pds-rings.seti.org/cassini/report/Cassini\%20Final\%20Report\%20-\%20Volume\%201.pdf} \citep{FinalReport}.

\hspace{1cm} \textbf{How old are Saturn’s main rings? Are young rings consistent with our understanding of the formation and evolution of the Saturn system as a whole?} Determining the mass of Saturn’s main rings was a principal measurement goal of Cassini’s proximal orbits, the final 22 orbits where Cassini flew between the D ring and Saturn’s cloud tops. The mass of Saturn’s rings ($0.4 \, \textrm{M}_{Mimas}$), derived from gravity measurements \citep{Iess2019} and localized surface mass density determinations \citep{Spilker2004, Colwell2009, Tiscareno2013, Hedman2014, Hedman2016}, has been interpreted, in consideration of the micrometeoroid mass influx measured by Cassini’s CDA \citep{Altobelli2016, Kempf2017}, to imply a young age for the rings, $\sim10^7 - 10^8$ years. A recent origin for a ring system as massive and extensive as Saturn’s holds implications for the rest of the system. Dynamical scenarios consistent with a young ring predict a relative abundance of impact craters among the inner moons of Saturn generated by planetocentric impactors \citep{Cuk2016}. Recent findings suggest an enhancement in the small-crater population \citep{riveravalentin2018, ferguson2019} that may be consistent with such a scenario and is an active area of research.  Further discussion on the determination of the age of Saturn’s rings and its significance can be found in \citep{Hsu2020}.

\hspace{1cm} \textbf{What does the pollutant (non-ice) material detected in Saturn’s rings indicate about the physical and chemical history of Saturn’s environment?} Cassini provided insight into the composition of Saturn’s rings, but details remain stubbornly murky. VIMS near-infrared observations reveal correlations between the rings’ non-ice fraction and the rings’ structure. Based on these near-IR data \citep{Nicholson2008, Hedman2013} and on ultraviolet observations \citep{Bradley2010, Bradley2013}, two distinct contaminants have been identified: a broadband absorber localized in the C ring and Cassini Division and a broadly-distributed contaminant that absorbs at short visible and UV wavelengths. Cassini RADAR observations suggest silicate concentrations of 6-11\% by volume in the mid-C ring, some 4-10\% higher than elsewhere in the C ring \citep{Zhang2017}. Further out, Saturn’s E ring is believed to be fed by cryovolcanic plumes of contaminated water ice emanating from near Enceladus' south pole \citep{Spahn2006}; these particles likely contribute to the surface contamination of the midsize icy moons \citep{Schenk2011}. Contamination fraction is a key piece in understanding the rings’ exposure age. Cassini’s proximal orbits also permitted the in situ measurement of Saturnward-drifting material sourced from the rings. Spatial distribution and compositional information was obtained with Cassini’s RPWS \citep{Ye2018}, CDA \citep{Hsu2018} and INMS \citep{Waite2018} instruments. INMS data conclusively reveal the presence of $ \textrm{CH}_4, \textrm{CO}_2, \textrm{CO}, \textrm{N}_{2}, \textrm{H}_{2}\textrm{O}, \textrm{NH}_{3} $ and organics \citep{Miller2020}. CDA results include the identification of silicate material in nanograins \citep{Hsu2018}. So why is the spectral evidence for any of these species inconclusive at best? Additional laboratory work may provide the answer.

\begin{wrapfigure}{R}{0.6\textwidth}
\centering

    \includegraphics[width=.6\textwidth]{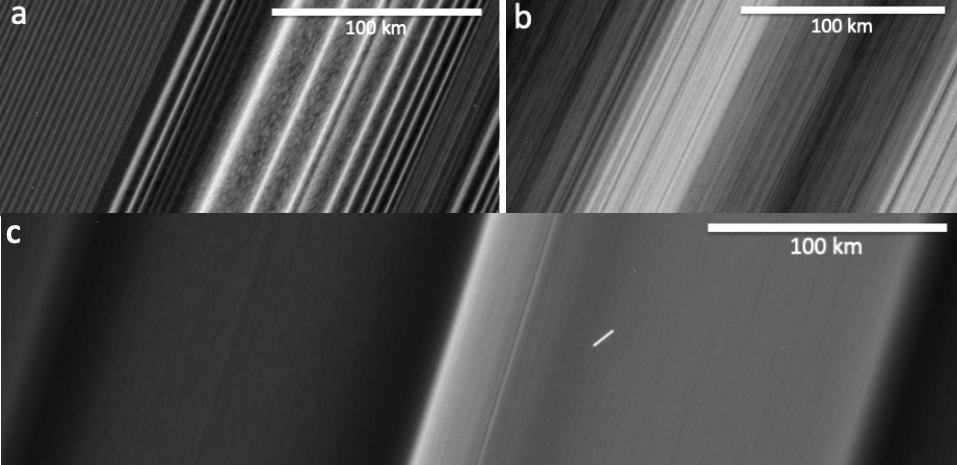}
    \caption{Cassini images illustrating medium-scale structure discovered in Saturn’s rings, including density waves triggered by satellite resonances (a), unexplained structure in the B ring (b), and the innermost C ring plateau (c). \textit{These features highlight the complex dynamical interaction between ring particles and external gravitational forces.} }
    \vspace{-20pt}
    \label{fig:meoscale}
\end{wrapfigure}

\hspace{1cm} \textbf{Why does Saturn's interior have such complex structure, and what does this tell us about giant planets in general?} Cassini has shown how rings can probe planetary interiors, complementing gravity and magnetic field data. The spiral density and bending waves used to retrieve information such as local surface mass density and viscosity, are typically raised by distant satellites. But occultation profiles reveal waves in the C ring raised by internal oscillations and gravitational irregularities \textit{within Saturn itself} \citep{Colwell2009, Baillie2011, Hedman2013b, Hedman2014, French2016, French2019, Mankovich2019}. Such waves were first identified in Voyager radio occutation data \citep{Rosen1991}. Combining the analysis of these features with analysis of gravity and magnetic field data may yield new information about the structure of Saturn’s interior and provide another measure of the planet’s rotation rate. It may well be a viable measurement technique at other ringed bodies.

\hspace{1cm} \textbf{How do the complex structures in Saturn's rings form?} Cassini images and occultations reveal structural features at a wide range of spatial scales, from localized C ring “ghosts” \citep{Baillie2013} to the Cassini Division itself.  Such structures include non-axisymmetric clumps or density enhancements in the local distribution of ring particles. Accretionary processes likely play a role in the formation of some if not all of them. Propellers, potential “missing links” to the progenitors of the rings, have been observed at multiple locations \citep{Spahn2018, Tiscareno2006, Tiscareno2008}. Further study of their size and spatial distribution may provide insight into the origins of Saturn’s rings. We still do not fully understand how spokes form. A large number of ring structures remain to be explained.\medskip

\textbf{\textit{Jupiter}: }
Jupiter’s dusty ring system remains the only one discovered by spacecraft. Hints of an unseen ring from Pioneer 11 magnetometer data \citep{Acuna1976} were validated by Voyager 1 with a single, multiply-exposed image of the main ring \citep{Smith1979a}, with Voyager 2 revealing the ring system in detail \citep{Smith1979b}. The Jovian ring system has been observed by more spacecraft, including Galileo \citep{OckertBell1999}, Cassini \citep{Throop2004}, New Horizons \citep{Showalter2007} and Juno \citep{Becker2019,Ye2020}, than any other.

\hspace{1cm} The rings of Jupiter are characterized by their low optical depth and the non-gravitational forces that sculpt them. The main ring extends from $ 1.72 - 1.81 \, \textrm{R}_{J} $ \citep{OckertBell1999}. Enhanced backscatter suggests a narrow belt of macroscopic particles, the purported dust source of the main ring and the vertically extended halo interior to it, orbiting between Metis and Adrastea \citep{Burns1984,Showalter1987}. Galileo and Keck images \citep{OckertBell1999,dePater1999} reveal two components to the gossamer rings, whose outer edges coincide with the orbits of Amalthea and Thebe. Jupiter’s rings are surprisingly dynamic and shaped by interactions with its magnetospheric environment that we are just beginning to understand.

\hspace{1cm} \textbf{What are the principal mechanisms of radial particle transport and how do they vary across the jovian ring system?} The \citet{Burns1999} model for the Amalthea and Thebe gossamer rings appeals to Poynting-Robertson drag to drive the inward migration of impact-derived dust. But outward extensions to these rings suggest a more complicated picture \citep{Showalter2011}. Photometric models of the main ring \citep{McMuldroch2000,Brooks2004,Throop2004} suggest a relative dearth of particles larger than $\sim 10 - 20 \, \mu \text{m} $. Is this the result of particle evolution and dynamics? How main ring particles migrate inward to form the halo and their subsequent evolution is equally unclear. Properties that determine halo particles’ charge-to-mass ratio, such as their susceptibility to photocharging and the local electrostatic potential, dictate their coupling to the magnetosphere and consequent dynamics and are poorly known. Lorentz resonances in the halo likely play a role \citep{Schaffer1992, Hamilton1994, Jontof-Hutter2012a, Jontof-Hutter2012b}, but alternative particle transport mechanisms have been proposed \citep{Horanyi1996, Horanyi2010}. The spatial distribution of dust impacts detected by Juno’s Waves instrument \citep{Ye2020} recalls the original inner disk \citep{Jewitt1981} that was later shown not to exist \citep{Showalter1987}. Where do these dust measurements fit in the broader picture?

\hspace{1cm} \textbf{What is the origin of the fine structures within Jupiter’s rings? What can they tell us about the environment at Jupiter?} For such an optically thin ring system, Jupiter’s rings contain an unexpected amount of fine-scale structure. Researchers have commented on a broad near arm/far arm asymmetry in Voyager \citep{Jewitt1981,Showalter1987} and Galileo \citep{OckertBell1999,Brooks2004} data, which is notably absent in Cassini \citep{Throop2004} and New Horizons \citep{Showalter2007} data. Vertical corrugations in the main ring imaged by Galileo \citep{OckertBell1999} (note the main ring's subtle, alternating bright and dark patches in the cover image) have been attributed to a disturbance somehow induced by the Comet Shoemaker-Levy 9 Jupiter impact \citep{Showalter2011}. This mechanism has been invoked in the more optically thick C and D rings of Saturn \citep{Hedman2007b,Hedman2011}. Clumps in the main ring are another example of fine-scale structure \citep{Showalter2007}. Finally, \citet{Showalter2008} note an enhancement of dust just interior to Amalthea and Thebe, suggesting, at least in Amalthea's case, that ring material is hung up in that satellite’s Lagrange points. The short lifetimes of dust at Jupiter imply that these features are either actively maintained or are ephemeral, caught at just the right time by visiting spacecraft.\medskip

\textbf{\textit{Uranus}:} The dense, narrow Uranian rings are immersed in a sea of micron-sized dust grains, which is strikingly different from Saturn's massive ring system, Jupiter's ephemeral rings, or the dusty rings around Neptune. Many open questions regarding their origin and evolution remain.

\hspace{1cm} \textbf{What processes define the structure — including the locations, widths, eccentricities, and inclinations — of the Uranian rings?} The small Uranian satellites likely play a significant role in forming, sourcing, and sculpting the ring system, though none of these interactive processes are well understood. The confined, narrow structure of the rings suggests unseen shepherding satellites, though more dynamically complex confinement mechanisms are required to explain  Saturn’s narrow F ring \citep{Cuzzi2014}. Currently, there is no definitive explanation for the spacing between the narrow rings, though it is known that these dense rings oscillate on orbital timescales against the backdrop of a much more slowly evolving (over several decades) dusty ring system. A hypothesized, self-sustaining mechanism for narrow rings has not been confirmed in detail; refinements would likely apply to other narrow ring systems, such as those of Chariklo \citep{Nicholson2018}.  The satellite Mab \citep{Showalter2006} is embedded in, and is presumably the direct source for, the dusty $\mu$ ring.  Further study could elucidate the moon’s environment, including its meteoritic bombardment rate.

\hspace{1cm} \textbf{What is the origin of the Uranian ring system and how has it evolved?} What little is known about the composition of Uranus’ larger moons is intriguing: though H$_2$O rich, CO$_2$ ice appears preferentially on their trailing hemispheres \citep{Cartwright2015}, while the near-IR spectra of the rings and smaller satellites remain featureless \citep{deKleer2013}. This suggests that their composition may depend upon their present environment as well as their origins. If the rings are remnants of the original source material for the planet, their composition may be indicative of the formation and migration history of Uranus, especially when compared with the composition of Neptune’s rings. 

\hspace{1cm} \textbf{What can we learn about Uranus from ring studies?} Satellites and rings capture critical details about the magnetosphere and interior of their host planet; observations of radiolytic processing on the satellites and rings, as well as searches for ring rain along magnetic field lines as seen at Saturn \citep{Connerney1984,Waite2018} would provide insight into the tilted magnetosphere. Oscillations within the planet that trigger waves in the rings can be used to probe Uranus' interior \citep{Hedman2009, Hedman2013}.  Notably, ring precession rates are still the best constraints on Uranus' $J_2$ and $J_4$ gravity harmonics \citep{French1988}. \medskip

\textbf{\textit{Neptune}: } The Neptunian ring system consists primarily of micron-sized dust, which has a relatively short lifetime and must be replenished constantly. Denser arcs of material embedded within the outer Adams ring have been observed to change in brightness, drift in position, and even completely vanish \citep{dePater2018}. Smaller scale changes and compositional detail about the ring system are difficult to assess due to the low optical depth of the rings and Neptune's distance.

\hspace{1cm} \textbf{What is the origin of Neptune's rings system?} The rings and small moons may be remnants of the original Neptune system that was disrupted during the capture of Triton, and therefore may provide constraints on the inventory of material at Neptune's initial orbital distance from the Sun \citep{Goldreich1989, Banfield1992}. Better compositional measurements of the rings and satellites could determine if they share a common origin. Comparisons with Triton’s composition would elucidate differences in the primordial material of the Kuiper Belt versus the region in which Neptune formed. Finally, better constraints on the particle size distribution of the rings would be indicative of the collisional age of the rings. 

\hspace{1cm} \textbf{How are the Neptunian rings and arcs sustained?} The arcs’ stability and confinement are still areas of active research, with solar radiation forces and inelastic particle collisions challenging their maintenance through resonances or co-orbital moonlets (\citep{Foryta1996, Hanninen1997, Namouni2002, dePater2018}). Resonances may be used to explain the location of the LeVerrier ring, though whether that resonance is with an undiscovered moon or driven by structure within the planet is unknown. The short timescale changes that have been observed from Earth \citep{dePater2018} suggest active evolution of the rings and arcs, perhaps similar to what is observed at Saturn’s F ring \citep{Murray2008}. 

\hspace{1cm} Intriguingly, the radial layout of the rings and moons at Neptune are the reverse of that at Uranus. The innermost moons of Neptune are effectively as close to the planet as the Uranian rings, so understanding how Neptune’s inner moons hold themselves together could explain why these ring systems are so different. Further, unlike the Saturnian ring satellites, most of Neptune’s small satellites reside inside synchronous orbit. This means that the moons are moving inward until they become tidally disrupted, with unique consequences for the structure, development, and possible recycling of the Neptunian rings and moons. The capture of Triton likely had a significant dynamical effect on the Neptune system, most evident by the highly eccentric satellite Nereid. Investigating how the rings responded to this disruption event would provide unique insights into the evolution of planetary disks facing large-scale disturbances. 

\vspace{-10pt}
\section*{\large New \& Old Ring Discoveries}
\vspace{-10pt}
An entirely new regime of rings science was uncovered within the last decade when discrete rings were discovered around the Centaur Chariklo \citep{Braga-Ribas2014}, the Kuiper Belt Object Haumea \citep{Ortiz2017} and possibly around the Centaur Chiron \citep{Ortiz2015}.

\hspace{1cm} \textbf{How do rings form and evolve around small bodies, and how do they compare with the discrete rings of the outer planets?} Markedly distinct from scenarios invoked for ring formation around the outer planets, the existence of rings around small bodies suggests a new set of ring-forming mechanisms, including outburst activity or micro- or macro-impacts onto the primary body in addition to satellite breakups. Studying the structure and composition of the rings provide clues to their origin. Of paramount interest is how water-ice-rich rings could be generated around Chariklo, though the central body lacks any H$_2$O spectral features \citep{Duffard2014}. This paradox could have strong implications for how the rings formed, including the presence of subsurface ice if the rings were created from impacts or activity from the Centaur.

\hspace{1cm} \textbf{How stable are rings, and can they constrain the dynamical evolution of small bodies?} Centaurs are presumably primordial objects, possibly originating from the Kuiper Belt, that have been perturbed into planet-crossing orbits with finite lifetimes of $\sim 10^{6}$ years. The rings’ structure and the potential presence of smaller satellites to generate and/or gravitationally sculpt the rings may constrain the evolution of the system, guiding interpretations of its age. The age and stability of the rings have implications for previous dynamical interactions with the outer planets as the Centaurs' orbits evolved, and may be used to trace their movement through the Solar System.

\hspace{1cm} \textbf{Did other Solar System objects once host rings?} Intriguing numerical simulations suggest that Phobos and Deimos may have coalesced from an ancient ring that dissipated through deposition of debris onto Mars in a cyclical ring-moon formation process \citep{Hesselbrock2017}. Observational evidence of a potentially collapsed ring exists in the form of an equatorial ridge on Saturn’s moon Iapetus \citep{Stickle2017}. A more complete understanding of the evolution of ring systems may be gleaned through the establishment of the end-state of ancient rings on various planetary bodies \citep{Sicardy2018}. 

\hspace{1cm} \textbf{In what type of environment could a ringed exoplanet have evolved?} With more sophisticated instrumentation, rings will be detected around exoplanets. Applying our knowledge of the diverse ring systems in our own Solar System to any structural or compositional measurements of exo-rings will result in substantial advances in understanding the origin and evolution of ringed exoworlds and their host environments.

\vspace{-10pt}
\section*{\large Key Recommendations for NASA}
\vspace{-5pt}
\textbf{We strongly urge NASA to continue to maintain a robust Research \& Analysis program to support planetary rings science.}
\vspace{-5pt}
\begin{itemize}
  \item The Cassini mission resulted in a deep trove of data of Saturn’s rings. Continued analysis, including cross-instrument comparative science, is still needed. We urge continued support for the CDAP program, as well other data analysis programs (e.g., NFDAP) that can support rings science, well into the next decade. Finally, we recommend continued support for the SSO to conduct observations from Earth-based facilities.
  \vspace{-5pt}
  \item Observational data of rings can only be interpreted in the context of ground-truth data. We urge an increase in funding for laboratory studies that make relevant compositional measurements, conduct ring particle interaction experiments, and theoretical modeling projects through R\&A programs like SSW. 
\end{itemize}

\medskip
\textbf{We strongly endorse NASA support for Earth-based rings science observations.}
\vspace{-5pt}
\begin{itemize}
  \item Stellar occultation campaigns are the leading technique for discovering new ring systems \citep{Elliot1977,Hubbard1986,Braga-Ribas2014,Ortiz2017}, enable detailed measurements of ring structure and particle size distributions \citep{French2000}, and provide hazard mitigation support for NASA missions (\citep{Throop2015}). We urge support for new and existing facilities with the necessary time resolution for stellar occultation measurements, as well as campaign efforts using smaller, portable telescopes that can involve collaborations between experts and non-experts (e.g., for Arrokoth \citep{Buie2020}). 
  
  \vspace{-5pt}
  \item Facilities covering broad ranges of wavelengths, from ground-based radar to space-based UV capabilities, are needed for the continued assessment of ring composition. High-resolution images enable deep searches for new rings and long-term monitoring for ring variations that answer some of the outstanding dynamical questions presented above, through, for example, a dedicated Solar System space-based telescope \citep{Young2020}.

\end{itemize}

\medskip

\textbf{We urge NASA to prioritize strong rings science goals when evaluating mission proposals to the outer solar system.}
\vspace{-5pt}
\begin{itemize}
  \item Planetary ring systems are being increasingly appreciated for what they tell us about their environment, the origins of their systems and the interior structure of the bodies they circle, as well as their relevance to circumstellar disks beyond our Solar System. Missions lacking rings science goals forfeit insights into the broader scientific pictures of these systems.
  \vspace{-5pt}
  \item Spacecraft provide unique platforms from which many of the open questions above can be answered. The value of in situ measurements to rings science has been definitively shown, most recently by Cassini and Juno.  Remote sensing observations benefit from geometries that cannot be obtained from Earth. New technologies will permit previously unattainable measurement goals.  For example, \citet{Hinson2017} describe a novel radio occultation experiment with New Horizons with sensitivity sufficient to penetrate even Saturn's B ring.
\end{itemize}

%\begin{figure}[!tbp]
%  \begin{center}
%  \includegraphics[width=0.9999\textwidth]{refs.png}

%\end{center}
%\end{figure}

\pagebreak
\begin{flushleft}
%\begin{multicols}{2}
\raggedright

%\end{multicols}
\end{flushleft}
\end{document}